\documentclass[conference]{IEEEtran}
\IEEEoverridecommandlockouts
\usepackage{cite}
\usepackage{amsmath,amssymb,amsfonts}
\usepackage{algorithmic}
\usepackage{graphicx}
\usepackage{textcomp}
\usepackage{xcolor}
\usepackage{color,soul}
\usepackage{hyperref}
\usepackage{enumitem}
\def\BibTeX{{\rm B\kern-.05em{\sc i\kern-.025em b}\kern-.08em
    T\kern-.1667em\lower.7ex\hbox{E}\kern-.125emX}}
\begin{document}

\title{Generative AI in Computing Education: Perspectives of Students and Instructors}

\author{\IEEEauthorblockN{Cynthia Zastudil}
\IEEEauthorblockA{\textit{Computer \& Information Sciences} \\
\textit{Temple University}\\
Philadelphia, PA \\
\href{mailto:cynthia.zastudil@temple.edu}{cynthia.zastudil@temple.edu}}
\and
\IEEEauthorblockN{Magdalena Rogalska}
\IEEEauthorblockA{\textit{Computer \& Information Sciences} \\
\textit{Temple University}\\
Philadelphia, PA \\
\href{mailto:m.rogalska@temple.edu}{m.rogalska@temple.edu}}
\and
\IEEEauthorblockN{Christine Kapp}
\IEEEauthorblockA{\textit{Computer \& Information Sciences} \\
\textit{Temple University}\\
Philadelphia, PA\\
\href{mailto:christine.kapp@temple.edu}{christine.kapp@temple.edu}}
\and
\IEEEauthorblockN{Jennifer Vaughn}
\IEEEauthorblockA{\textit{Computer \& Information Sciences} \\
\textit{Temple University}\\
Philadelphia, PA\\
\href{mailto:jennifer.vaughn@temple.edu}{jennifer.vaughn@temple.edu}}
\and
\IEEEauthorblockN{Stephen MacNeil}
\IEEEauthorblockA{\textit{Computer \& Information Sciences} \\
\textit{Temple University}\\
Philadelphia, PA \\
\href{mailto:stephen.macneil@temple.edu}{stephen.macneil@temple.edu}}
}

\maketitle

\begin{abstract}
Generative models are now capable of producing natural language text that is, in some cases, comparable in quality to the text produced by people. In the computing education context, these models are being used to generate code, code explanations, and programming exercises. The rapid adoption of these models has prompted multiple position papers and workshops which discuss the implications of these models for computing education, both positive and negative. This paper presents results from a series of semi-structured interviews with 12 students and 6 instructors about their awareness, experiences, and preferences regarding the use of tools powered by generative AI in computing classrooms. The results suggest that Generative AI (GAI) tools will play an increasingly significant role in computing education. However, students and instructors also raised numerous concerns about how these models should be integrated to best support the needs and learning goals of students. We also identified interesting tensions and alignments that emerged between how instructors and students prefer to engage with these models. We discuss these results and provide recommendations related to curriculum development, assessment methods, and pedagogical practice. As GAI tools become increasingly prevalent, it's important to understand educational stakeholders' preferences and values to ensure that these tools can be used for good and that potential harms can be mitigated.
\end{abstract}


\section{Introduction}

The introduction of Generative AI (GAI) models has led to significant excitement in the computing education community~\cite{sarsa2022automatic, macneil2022automatically, wermelinger2023github, finnie2022robots, macneil2022generating, macneil2023experiences, becker2023programming, savelka2023large, leinonen2023comparing, tran2023using}. These generative models are typically powered by large language models such as GPT-3 or Codex which are capable of producing code explanations~\cite{macneil2023experiences, leinonen2023comparing, macneil2022generating} that have been rated as being better than peer-generated explanations~\cite{leinonen2023comparing}. In addition, these models can solve basic programming assignments~\cite{sarsa2022automatic, becker2023programming} and perform slightly worse than students on quizzes with multiple-choice questions~\cite{savelka2023large}.
Understandably, these substantial and rapid advances in the performance of generative models are causing excitement and consternation among students and instructors. A recent birds of a feather discussion at the SIGCSE conference~\cite{macneil2023implications} highlighted a few emerging concerns that educators have related to over-reliance, plagiarism, and other forms of misuse. These concerns have been echoed by position papers~\cite{becker2023programming, kasneci2023chatgpt}, a working group~\cite{prather2023transformed}, and an investigation into instructor perspectives~\cite{lau2023ban}. However, these concerns only scratch the surface,
there is an urgent need to understand students' and instructors' preferences about and what they want from these GAI tools so that future development and regulation can align with the values of these critical stakeholders.

In this paper, we conduct an interview study with 18 participants including both students (N=12) and instructors (N=6). The goal of this interview study is to better understand these two key stakeholders' awareness of these models and their preferences about how these models should be used in computing education classrooms. We are particularly interested in what ways students and instructors envision the involvement of generative AI in computing education, both positively and negatively.

In this paper, we investigate three research questions:
\begin{itemize}[label={}]
    \item \textbf{RQ1}: What preferences do instructors and students have regarding how Generative AI should be used in computing education classes?
    \item \textbf{RQ2}: What concerns do instructors and students have regarding the use of Generative AI in computing education?
    \item \textbf{RQ3}: What implications might Generative AI have on computing education curricula, pedagogy, and assessment methods?
\end{itemize}

The results from our study suggest that students and instructors believe computing course curricula and assessment methods should be updated to include GAI tools; however, there are important challenges which need to be addressed in order to use them successfully while reducing their potential harm to instructors and students alike. These results provide a timely and important snapshot of students' and instructors' perspectives and preferences which can inform the design of pedagogies, tools, and policies that best align with students' preferences and needs.

\section{Related Work}

\subsection{An AI Phase-Shift in the Classroom}

Less than three decades ago, widespread access to the Internet fundamentally changed the education landscape. As internet access in the United States rose from 14 to 77 percent, students could suddenly access educational resources, play educational games, and interact with peers in online learning communities.
This inequitable access to these resources also led to a digital divide where some students were left behind because they did not have personal computers or reliable internet access~\cite{margolis2008stuck}. Recent advances in AI are poised to create a similar phase shift in the education landscape. AI-powered tools are becoming ubiquitous both in online and face-to-face learning environments. From a student's perspective, these AI-powered tools provide a number of important benefits such as an increased awareness of their performance, personalized learning pathways~\cite{basham2016operationalized}, and personalized support in the form of intelligent tutors~\cite{crow2018intelligent, anderson1985intelligent}. From an instructor's perspective, learning analytics tools provide instructors with the ability to monitor student activity and adapt their teaching~\cite{leitner2017learning, rienties2020defining, ihantola2015educational, cerratto2021analytics}. Some of these tools are even integrated into classroom environments where instructors can use augmented reality to monitor student learning, metacognition, and behavior in real-time~\cite{holstein2018student}. Chen et al. discuss additional ways in which AI in the classroom can be beneficial for instructors such as in improving student reviews, grading, and feedback, utilizing intelligent tutoring systems, and improving pedagogical practices and student experiences using AI-powered VR for experiential or practical learning experiences~\cite{chen2020artificial}.

However, many of these advances have been limited by the current abilities of machine learning. Recent advances in machine learning have introduced generative models capable of understanding and generating code have crucial implications for computing education. Researchers are already demonstrating the capabilities of these models to help students generate code~\cite{Puryear2022github, wermelinger2023github, denny2022conversing}, explain code to students ~\cite{macneil2023experiences, macneil2022generating, leinonen2023comparing}, and even produce programming assignments~\cite{finnie2022robots,sarsa2022automatic}. However, there are also some emerging challenges related to over-reliance and plagiarism~\cite{macneil2022generating, macneil2023implications}. For instance, a recent study showed how these models can perform nearly as well as students on multiple-choice questions~\cite{savelka2023large}.
So while a phase shift appears to be underway, it is a critical moment to ensure that things are changing for the better and not for the worse. 

\subsection{Carefully Considering the Design of AI Systems}

The introduction of AI systems has led to new design challenges related to their probabilistic nature, their lack of transparency, and their issues related to fairness and accountability. These challenges have led to rapidly evolving research topics such as explainable AI (XAI), human-AI interaction (HAII), and human-centered machine learning.  They have also led to the creation of new conferences that focus on fairness, accountability, and transparency. Across these fields, there is a resounding call to focus on approaches that are human-centered~\cite{riedl2019human}. For instance, XAI has grappled with questions like \textit{explainable to who and for what purpose}~\cite{ehsan2020human}. Additionally, people calibrate their trust through interactions and experiences with the AI system~\cite{lee2004trust}; however, other factors such as age, culture, and personality can also affect trust in AI systems~\cite{hoff2015trust}. Finally, there can be a tendency to develop an over-reliance on AI systems which is, unfortunately, most common for low-expertise users~\cite{gaube2021ai}. Based on these threads of research, understanding stakeholders is crucial for developing AI systems that are comprehensible, trustworthy, and avoid excessive reliance. Conversely, there are many examples of AI systems that violate these principles and cause harm~\cite{mayer2020unintended}.

Numerous computing education researchers are already addressing these issues of explainability, trust, and over-reliance. For instance, researchers are investigating the importance of AI explainability in computing settings~\cite{fiok2022explainable}. Khosravi et al. developed a new framework for XAI to guide researchers and practitioners to develop AI-powered education (AIED) systems that are more trustworthy, explainable, and best aid in learners’ educational goals~\cite{khosravi2022explainableAI}. Their framework outlined 6 important factors to consider when building XAI: who the stakeholders are (e.g., instructors and students), what benefits stakeholders receive from interacting with the system, what potential issues stakeholders may encounter when using the system, how explanations are presented, what AI models are used, and the best ways to design user-friendly and effective AIED systems. Frameworks that exist to guide the development of effective AIED systems that mitigate potential harm to all stakeholders are useful. However, the usage of GAI tools in education, such as ChatGPT and GitHub Copilot, poses a unique problem, as they were not developed solely for an educational purpose and their inclusion in the classroom may have unknown effects. Many researchers have begun to explore the potential negative effects the inclusion of these models into education may bring, including bias and fairness of these models, over-reliance, explainability, and trust~\cite{becker2023programming, macneil2023implications}.

\section{Methodology}
To investigate our research questions about student and instructor perspectives, we conducted semi-structured interviews with 18 participants. This qualitative research method is often used to elicit participants’ perceptions, experiences, and values regarding a particular topic or technology~\cite{peters2015interviews}. The field of computing education has begun adopting interview studies ~\cite{heckman2021literature, sheard2009analysis, lishinski2016rigor, schulz2022students}, often to explore the perspectives of instructors or students with respect to topics such as instructor pain points~\cite{mirhosseini2023pain} or student perspectives about online versus on-site teaching~\cite{jonsson2022student}. 
In this study, we focus on the perspectives of both instructors and students because it is critical to include all relevant stakeholders when considering the implications of new technology such as generative AI.

\subsection{Participant Demographics}
We conducted interviews with students (N=12) and instructors (N=6) which resulted in 18 total interviews. Participants were recruited from the Department of Computer and Information Sciences at a large R1 University in the United States. The instructors included a mix of tenure-track and teaching-track faculty. Students were recruited through student organizations, flyers, and snowball sampling. Student participants majored in either Computer Science (11/12) or Information Science and Technology (1/12). All of the instructors interviewed taught computer science courses.

\subsection{Interview Protocol and Analysis}
We conducted 30-minute semi-structured interviews. 
Interviews were semi-structured so that participants could partially guide the conversation toward insights they believed were important and allowed us to follow up on interesting aspects of the conversation~\cite{magaldi_semi-structured_2020}. The main questions from the semi-structured interview are included in the Appendix; however, we often asked additional follow-up questions to obtain further insights and clarification from participants.
Interviews were recorded and we thematically analyzed the interviews~\cite{braun2006thematic} by reviewing the transcripts, coding participants’ insights, and then identifying themes. Two researchers followed this process independently, then compared codes in order to help mitigate interpretation biases. Once the thematic analysis was completed for all of the interviews, both researchers compared and combined themes through mediation.

\begin{figure*}
\centering
  \includegraphics[width=0.8\textwidth]{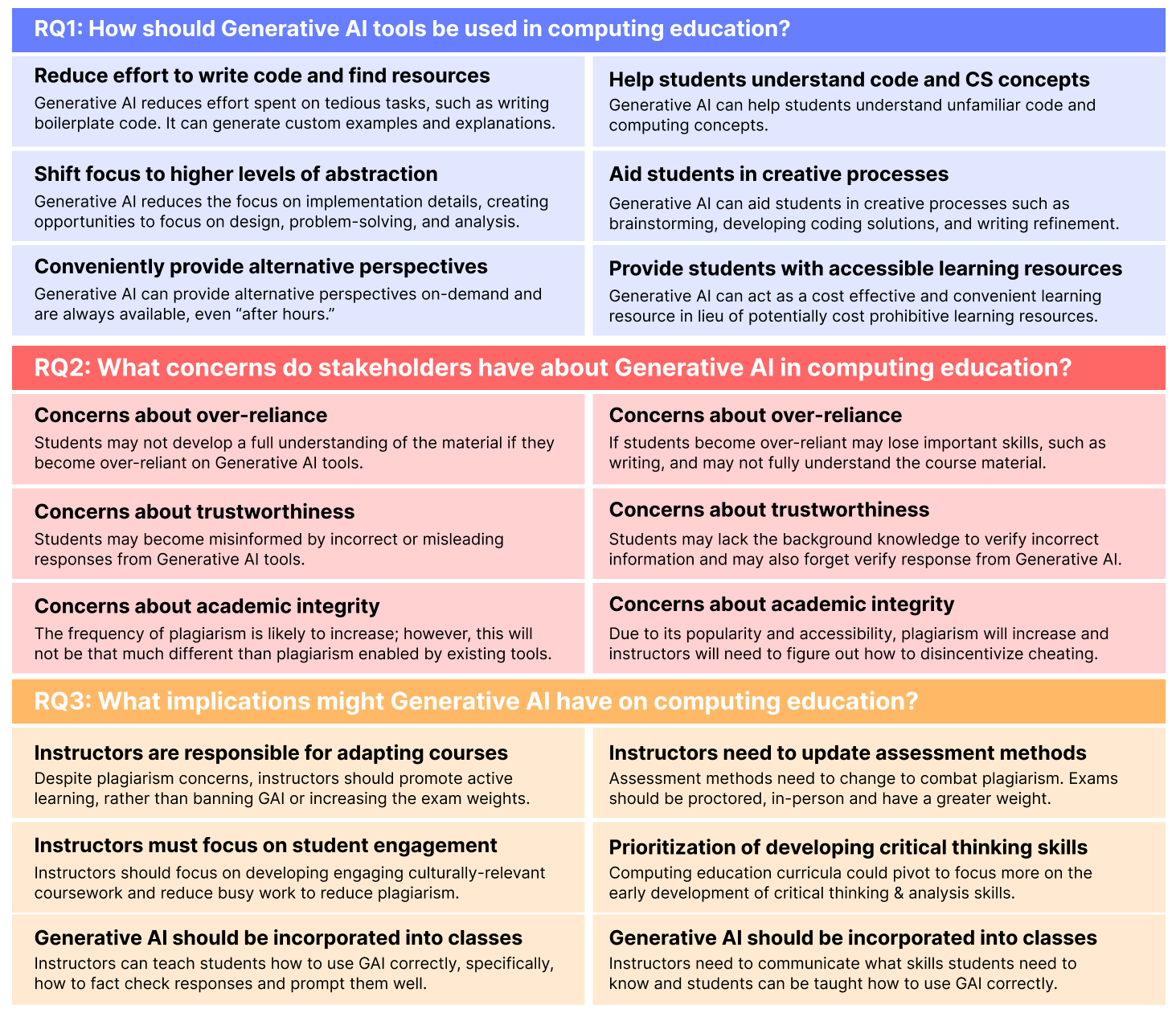}
  \caption{On the left, summarized themes from student interviews are presented and, on the right, summarized themes from instructor interviews are presented.}
  \label{fig:results}
\end{figure*}

\subsection{Limitations}

There is currently an urgent need to understand the implications of generative AI in computing education. For example, researchers and practitioners are already raising concerns about over-reliance, academic misconduct, and model trustworthiness~\cite{becker2023programming, kasneci2023chatgpt}. This work serves as an early probe into understanding current preferences, values, and concerns. While limited in scope, our work provides a preliminary investigation in order to help instructors, researchers, and more quickly adapt to this changing landscape.
There are a few limitations to our study which may impact the generalizability of our results and that warrant discussion. 
The participants of this study, both instructors and students, came from a single R1 university and represent a relatively small sample size, and as such, these results may not fully represent different demographics across different universities. However, prior work has shown that results obtained from qualitative studies with small sample sizes can still provide valuable insights for the topic being researched~\cite{zander2020copying, vanbeek2021feedback}.

\section{Results - Themes from Student Interviews}

In this section, we present the themes that emerged from our student interviews. A summarized version of the results from interviews with both stakeholders (students and instructors) can be found in Figure~\ref{fig:results}.Within each theme, we present sub-themes, our observations, relevant supporting quotes, and relationships between the theme and our three research questions presented in the introduction. Despite focusing our questions on Generative AI (GAI) tools such as GitHub Copilot, Codex, and Grammarly, our participants tended to focus on ChatGPT. 

\subsection{Perceived Benefits of GAI Tools (RQ1)}

Students shared the ways in which they have found GAI Tools to be beneficial in their computing courses, as well as their potential uses in computing education.

\textbf{GAI tools reduce the effort to write code and find learning materials.} Most students (10/12) believed that GAI tools, such as ChatGPT, have the potential to become, and in many cases are already, valuable learning tools to reduce the effort for students and instructors to complete some common programming tasks and find learning materials. Of the remaining two students, one was not very familiar with GAI tools and was therefore unsure of their value and the other actively tried to avoid using them.

Students shared a number of ways in which they have found GAI tools useful in the computing education context:
\begin{itemize}
    \item Generating skeleton or boilerplate code (S2, S3, S8, S9)
    \item Generating explanations of code (S2, S6, S7, S10, S11)
    \item Generating examples of concepts (e.g., linked-lists, sorting algorithms) (S2, S4, S5, S6, S11)
\end{itemize}
Reflecting on how GAI tools can generate examples to supplement existing learning materials, S2 said, 
\begin{quote}
\textit{``[Students] can use [ChatGPT] to find resources online... like, ‘Hey, I need a problem related to bubble sort,’ and then it will show them examples.''}
\end{quote}

\textbf{Students use GAI tools to avoid busy work.} When asked what motivates students to use these tools, students who frequently use them expressed that they were more likely to use GAI tools for assisting with, or even completing, assignments that they perceived as ``meaningless” or mere ``busy work.” Some students (4/12) compared it to existing tools, such as the calculator, whose introduction into the classroom allowed students to progress towards more advanced topics faster and reduce the amount of time students have to spend on familiar tasks or could be considered tedious~\cite{drijvers1996graphics}. According to some of the students, this refers to any work that takes too much of their time for very little academic benefit or that they aren’t passionate about. Participant S2 remarked on what kind of work they use ChatGPT for the most, saying,
\begin{quote}
\textit{``ChatGPT could just kind of do all the busy work or anything I could figure out, in like a couple of minutes. I’d rather ChatGPT do it, because it will.''}
\end{quote}

\textbf{GAI tools shift the focus to higher levels of abstraction.} Students expressed (6/12) that GAI tools could help reduce the amount of time spent working at lower levels of abstraction. Students often gave examples of how GAI tools could shift the learner's focus toward higher-level skills (e.g., design, the analysis and evaluation of code, or advanced software engineering concepts) which students are more likely to encounter in later computing courses.
For example, participant S9 described how students might focus on design patterns and software design rather than implementation details. They said,
\begin{quote}
\textit{``You don't usually get design patterns until you are in graduate school, right? Unless you'd specifically seek it out...% If 
we could use ChatGPT to help teach design patterns where...
You can use ChatGPT as a tool to implement all the little pieces. But we are going to try to break down this larger problem to figure out what pattern applies to it, and then use whatever you need to use to generate the individual pieces of code. But we really care about the overall structure and making sure that the students kind of have this broader understanding of the piece of software rather than a single algorithm or a single piece of code or a smaller scale program.''}
\end{quote}

\textbf{GAI tools provide alternative perspectives and sources of assistance.} Another common theme amongst students who regularly use GAI tools is that these tools provide another source of assistance outside and their instructors and teaching assistants (4/12). Some reasons why students said they choose to use a tool such as ChatGPT rather than asking their instructors or teaching assistants are:
\begin{itemize}
    \item Some professors are not good at explaining course concepts and code (S4, S5)
    \item GAI tools can provide multiple and alternative perspectives about programming concepts (S5, S10)
    \item GAI tools can be more convenient sources of assistance over instructors or TAs (S5, S8)
\end{itemize}
S4 described a time in which they wished they were able to use GAI tools in order to get more help with a concept in one of their computing courses, especially when they found themselves searching for a different perspective when they felt their instructor wasn’t communicating what they needed,
\begin{quote}
\textit{``With some professors, they don’t know how to communicate the answer that you’re looking for... I had a lot of trouble wrapping my head around it when I asked in what context will we use a linked list. [My Instructor] just explained what a linked list was, and with ChatGPT, I feel like I could say, ‘give me an example of a linked list and how I would implement it in this language,’ ''}
\end{quote}

\subsection{Concerns about GAI Tools (RQ2)}
While students largely expressed excitement about GAI tools’ benefits to computing education, students also expressed an amount of trepidation and concern regarding their use in computing education.

\textbf{Concerns about over-reliance.} Most students (9/12) mentioned the potential for students to develop an over-reliance on these tools as a concern. More specifically, students (6/12) detailed their concerns about the quality of education students receive if they become over-reliant on GAI tools that don’t adequately explain their responses. Some of the specific concerns about over-reliance and the quality of education were:
\begin{itemize}
    \item Students may not develop an understanding of the material when they get assistance from GAI tools (S1, S4, S5, S9, S11, S12)
\end{itemize}
The following quotes from participants further illustrate some of these student concerns. Participant S1 said,
\begin{quote}
    \textit{``When using them I don’t really fully understand what they’re telling me, and it’s just kind of like, ‘Oh! There’s the answer.’ But I’m not really the one learning and really digesting what’s happening and how I got to that conclusion.''}
\end{quote}
Participant S4 provided an additional perspective, saying, 
\begin{quote}
\textit{``I feel like a lot of people just heavily rely on it, and they use it to just copy and paste the work, and not necessarily understand why. I feel like they would just copy the work given to them. It kind of ruins the purpose of learning it in the first place.''}
\end{quote}

\textbf{Concerns about trustworthiness.} The majority of students (7/12) also expressed concerns about the trustworthiness of the output given by GAI tools. Students raised concerns about the reliability of information generated by these models due to their potential to misinform students.
Specifically, students outlined the following concerns about trustworthiness:
\begin{itemize}
    \item Information sources are missing or incorrect (S1, S8, S9)
    \item GAI tools tend to hallucinate information (S6, S8, S10)
    \item Students have to double-check everything, not just information, but things like false context and tone (S4, S6)
\end{itemize}
Participant S8 said, due to these models’ tendency to hallucinate students have to be careful to double-check the information they receive, 
\begin{quote}
\textit{``[ChatGPT] is helpful, but I understand that sometimes it also hallucinates as well. It will give you false information. I don't really know if it is yet like a Google replacement..., I feel like it's only as useful if you really know the topic that you're talking about, and like you can read what it’s saying, and be sure that it's not like [meaningless output], because sometimes you do have to search again on Google just to make sure what it said is correct.''}
\end{quote} 
Participant S4 also described needing to check responses for correctness, sharing that it wouldn’t stop them from using GAI tools altogether and it prevents them from relying on the tools too much,

\begin{quote}
\textit{``[Incorrect ChatGPT responses] wouldn’t keep me from using it again. It’s just a small detail. It keeps me on my toes, and makes sure I double-check everything. I can’t rely too much on it.''}
\end{quote}

\textbf{Concerns about academic integrity.} The majority of students (7/12) also brought up the issue of academic integrity. These students expressed the belief that the amount of plagiarism will increase as these tools become more popular.

However, students did not necessarily believe that AI-aided plagiarism is all that different from plagiarism that occurred previously. Students claimed that the only difference between plagiarizing by copying someone else’s work or paying for a subscription service such as Chegg and using a tool like ChatGPT is that it’s free and available to everyone.  
\begin{quote}
\textit{``Before I've seen people use things like, Chegg, to just straight up just finish their homework. I feel like it's just a difference between you using ChatGPT, which is a free service, and it can give you the information you want, and you could probably just change up a couple of things with it versus paying a fee for a subscription-based service that just hosts the work that your professors are giving to you.'' (S4)}
\end{quote}

\subsection{Predicted Implications of GAI Tools (RQ3)}
Students were excited to share their insight about how course curricula and assessment methodologies should be adapted to accommodate GAI tools.

\textbf{Instructors are responsible for adapting their courses.} In our interviews, none of the students said that instructors should ban the use of GAI tools. On the contrary, multiple students (5/12) expressed that instructors are responsible for adapting their courses in ways to work with this new technology, as instructors have had to in the past with any innovative and popular technology, such as calculators, Wikipedia, and the Internet (4/12). While the majority of students expressed that they were not sure exactly what steps need to be taken by instructors in order to adapt their classes, a couple of students indicated updates they hope instructors do \textit{not} make to their classes. Participants S3 and S8 specifically mentioned one way to avoid plagiarism would be to increase the number of exams or the weight of exams in the final grades of courses. However, both students said they would not want this,
\begin{quote}
    \textit{``I guess they would have to possibly lower lab weights and maybe increase exam weights, or you know things that can be more so tested in person... 
    %Which kind of sucks, because, like it's not, 
    I feel like it shouldn't be like that...  I kind of had the benefit of, like, you know, being able to use Google, having like a 40-50\% like lab weight. Being able to use Google, learn, read a lot of documentation, and you know, actually, kind of somewhat suffer through labs, but also I knew that the lab grade would save me because I wasn't as good on exams,'' (S8)}
\end{quote}

\textbf{Instructors must focus on student engagement.} Several students (3/12) specifically mentioned that, when learning new material, hands-on active-learning assignments are the most helpful and engaging. Students (4/12) emphasized that they believed that engaging courses and coursework would more likely result in engaged students who see value in and want to complete the assignments and are less likely to plagiarize. Participant S1 said the following,
\begin{quote}
    \textit{``Instructors need to understand why students use ChatGPT in the first place, and some of the reasons why they use it is they just don’t understand the content, and they want an easy out and that means that something is going wrong with the curriculum that's being presented to them.''}
\end{quote}

\textbf{Instructors should incorporate GAI tools into the classes.} Students were excited about the potential for instructors to include these models in their curricula. Students suggested that instructors could focus less on implementation details and more on high-level concepts. Participant S5 pointed to the importance of other skills that should be learned in early programming courses, particularly problem-solving.
\begin{quote}
    \textit{``I think learning the language is the easiest part, but I think the introductory programming courses offer something a little bit more important, which is problem-solving. Which I think will still ... be a very important part of programming, because that's really all programming is. It's just problem-solving... But the problem-solving mindset is definitely something that cannot just be assumed people have.''}
\end{quote}

Students (4/12) also suggested that instructors could teach students how to use GAI tools by teaching students the best ways to prompt tools such as ChatGPT and the limitations and risks of using these tools. A couple of students also suggested instructors could use GAI tools to scaffold learning, as described in more detail by participant S5, 

\begin{quote}
\textit{``It's like asking a mathematician to do mundane repetitive math equations today without a calculator... yeah, you could do it, but do you want to? Not really. So, I think instructors should, you know, start off teaching their students, you know, how to do it. How to do everything without an AI language model to help them. And then, as they get more advanced with it, they, you know, introduce the AI to just, you know, make the more difficult tasks a bit easier.''}
\end{quote}

\section{Results - Themes from Instructor Interviews}
The following section details the themes and observations discovered through our instructor interviews. Similar to students, instructors tended to focus primarily on ChatGPT.

\subsection{Perceived Benefits of GAI Tools (RQ1)}
While instructors (5/6) have not used GAI tools frequently, they envision a number of ways in which they can be beneficial for students in computing courses. Overall, instructors were familiar with GAI tools like ChatGPT and GitHub Copilot, but used them much less than students. Consequently, they had fewer insights about they might be used in the computing education context. However, every instructor (6/6) thought that these tools could be valuable for students.

\textbf{GAI tools can help students understand code and computing concepts.} A couple of instructors (2/6) thought that GAI tools were beneficial for students when they encounter unfamiliar code and that GAI tools could provide just-in-time support. While prior work has demonstrated the ability of generative AI to explain code, participant I2 described how this helps students when they encounter unfamiliar code online,
\begin{quote}
\textit{``You can give [ChatGPT] a little piece of a program you got from Github or whatever, and have it explain to you what this code is doing, and it would do a good job.''}
\end{quote}

\textbf{GAI tools can help students find inspiration, brainstorm, and get feedback on ideas.} A majority of the instructors (4/6) detailed ways in which GAI tools could be used to aid in creative processes such as ideation and brainstorming. For example, participant I1 described how GAI tools can help with brainstorming solutions to coding problems. Additionally, some instructors (2/6) described ways in which they've found success in having GAI tools, specifically ChatGPT, provide students with feedback on their writing for project-based computing courses. Participant I2 described how they've had students use ChatGPT to try to find weaknesses or areas of improvement for their essays or project proposals.

\textbf{GAI tools' low cost and accessibility can provide students with high quality learning resources.} A couple of instructors (2/6) expressed their excitement about how the low cost and barrier to entry can provide students with resources that can be prohibitively expensive, such as private tutors or paid services such as Chegg or CourseHero. Additionally, GAI tools could act as high quality learning resources outside of students' instructors and TAs, providing another perspective on course topics.

\subsection{Concerns about GAI Tools (RQ2)}
Instructors hold many common concerns regarding AI systems for the use of generative AI in computing education and their use by students.

\textbf{Concerns about trustworthiness.} Most of the instructors (5/6) cited the lack of trustworthiness of responses from GAI tools as one of their primary concerns. All of the instructors mentioned that it’s important to double-check the information received from these models. However, as one instructor pointed out, students don’t always have the necessary foundational knowledge to know whether the outputs are right or wrong. Participant I4 said, 
\begin{quote}
\textit{``We don't know that [ChatGPT] is giving [students] the right information. and you know, if I know a topic, I know when I'm getting bogus information. They don't, you know, so that's a concern.''}
\end{quote}

\textbf{Concerns about over-reliance.} Several instructors (3/6) also cited over-reliance as a significant concern when considering students using these models in their classes. Participant I3 expressed concern about students not fully learning material if they rely too heavily on GAI tools. Participants I1 and I3 both expressed concern that students may lose important skills, such as writing, if they become too reliant on these models. Participant I5 discussed their concern about GAI tools replacing students' own effort, saying,
\begin{quote}
    \textit{``I think pedagogically, we'll have to kind of go back to the drawing board. And come up with new ways to assess. Not necessarily just to prevent students from using and benefiting from these tools, by all means we should embrace change and new methods and new ways of doing things. So find ways where students can use these things to assist, to help but not one where it replaces their own cognitive load.''}
\end{quote}
\textbf{Concerns about academic integrity.} All instructors (6/6) expressed concern about students using GAI tools to plagiarize. Several instructors (4/6) expressed that, due to the availability and low cost of tools like ChatGPT, students are likely to plagiarize more. However, one instructor (I5) said,
\begin{quote}
    \textit{``I feel like a student who was never motivated to cheat in the first place will not do it now just because ChatGPT is available. Sufficiently motivated students would have always cheated. So if I were to lazily assume that there is an uptick in the number of students who are now exploiting this tool or an uptick in the number of students who are now cheating as a result of this tool, it would have been those in the middle who were always willing to, but perhaps lack the time, resources, access, or the know-how to properly engage in it, but I don't think just there's this giant movement towards more students doing it, who would not have done it otherwise. I don't think it's turning students into cheaters. I think it's just making it more accessible to those who already would have done it.''}
\end{quote}
This participant’s insight aligns with the perspectives shared by the other instructors in that, the concerns about academic integrity which have arisen due to the presence of GAI tools in the classroom, are less focused on just stopping students from cheating, which is unlikely, but rather in developing ways in which instructors can disincentivize cheating.

\subsection{Predicted Implications of GAI Tools (RQ3)}
Instructors shared uncertainty about what exactly needs to change in course designs and assessment methods; however, they emphasized the importance of proactively incorporating these models into computing education curricula.

\textbf{GAI tools should be incorporated into classrooms.} Unanimously, instructors (6/6) emphasized that they believe GAI tools should not be banned, and they should embrace their existence and include them in the classroom.
Instructors believed it was imperative to understand how these tools work. Some of the ways they envisioned incorporating GAI tools into their courses included:  
\begin{itemize}
    \item Let students use it but they should be able to explain why the output is correct or incorrect (I1, I3, I6)
    \item Clearly articulate what students need to know how to do without using GAI tools (I4)
\end{itemize}
Participant I4, detailed how, with any new tools which shift how courses are taught, clear expectations for students need to be established,
\begin{quote}
\textit{``... before calculators were a thing we used to expect students to memorize multiplication and division tables. And you know, after the calculator came into being, of course, you know it's reasonable not to expect people to memorize that when you know they've got a tool that they can use to figure out how to multiply things, you know. And I could see the same thing happening with us with programming now that tools like ChatGPT exist, you know, maybe there are certain things that we could expect students to use a tool for, and we don't have to test them on it. But at the same time, you know, there has to be some core level of knowledge that we should expect from a student, you know, without the use of tools.''}
\end{quote}

\textbf{GAI tools can facilitate critical thinking} A couple of instructors (2/6) were excited about the potential to use these models in their courses to help students further develop their critical thinking and analysis skills. Participant I6 said,
\begin{quote}
    \textit{``
    Let them actually use ChatGPT to come back with a report as a first draft. But then the job of the student is not simply to return the report, but instead, look at the report ChatGPT wrote, and then decide or criticize whether it's right or wrong, and then, if it's wrong, then what is wrong with it, or try to improve...
    So in that sense, it is more useful for the student to develop their critical thinking...
    instead of just like the old way of just writing a report.''}
\end{quote}
Additionally, participant I5 said,
\begin{quote}
    \textit{``Just the production of something is not going to be as important, perhaps leaning more into analysis, perhaps leaning more into your ability to troubleshoot. Versus creation which is one of which is one of the steps that you know of many steps in any sort of understanding or building a system or an algorithm or a data structure or anything else. ...It's not, give me code, it's, here is some code, give me some sort of analysis that matches or fits into something that is not easily template-tized, something that is not easily described or describable.''}
\end{quote}
\textbf{Instructors need to update assessment methods.} In terms of how to update curriculum to adapt courses to account for the existence of these models, instructors did not have a very clear idea of what exactly needed to be done, with many expressing uncertainty about the best way to assess student progress, prevent plagiarism, and ensure that students do not rely too heavily on the models, hindering their education. In their interviews, instructors did mention a few considerations they had for updating their assessment methods:
\begin{itemize}
    \item Mitigate plagiarism by developing engaging assignments and reducing busy work (I3)
    \item Change course grade weights such that easily plagiarized assignments can’t carry students’ grades (I4, I5)
    \item Continue to give proctored exams since they can’t be plagiarized like other out out-of-class assignments (I4)
\end{itemize}

\section{Discussion \& Conclusion}

\subsection{Converging Themes across Stakeholders and Opportunities}

In this section, we present converging themes that emerged through our stakeholder interviews. Based on these areas of convergence, we present some opportunities to maximize the benefits of generative models in computing education settings.

\subsubsection{Students and instructors are concerned about trustworthiness and over-reliance}

Across most interviews, students and instructors were concerned about the trustworthiness of GAI tools as well as the potential for students to become over-reliant on them. Participants highlighted the black-box nature of GAI and their tendency to hallucinate non-factual information. This raised concerns about students not understanding the responses they receive from GAI, a lack of reliability of the information received, and the impact of over-reliance on students’ learning outcomes.

\underline{\textit{Opportunity \#1:}} Explicitly teach students when and how to use GAI tools.
Educators can teach courses and modules about
how to fact-check the model's output in cases where the model is known to perform poorly; calibrating a healthy skepticism in generative AI. Additionally, instructors should not assume computing students have AI literacy. It may be necessary to teach courses in computational thinking and AI literacy~\cite{long2020literacy}.

\subsubsection{Students and instructors believe GAI tools will broaden access}

Students expressed that they believed tools like ChatGPT are valuable learning tools, partially due to the fact that it’s available at any time and it’s free or inexpensive to use. 
\underline{\textit{Opportunity \#2:}} GAI tools democratize access to help. %high-quality educational tools and level the playing field with paid education services. 
Similar to when unequal access to the Internet caused some students to be left behind~\cite{margolis2008stuck}, today, unequal financial access to tutors, unpaid internships, college prep programs, and paid content can have similar effects. GAI tools may offer cheaper access to high-quality tutoring. 

\subsection{Diverging Themes across Stakeholders and Challenges}
In this section, we present some diverging themes which have the potential to cause tension between students and instructors. We also describe how these divergences in preferences may eventually lead to challenges.

\subsubsection{Students tended to be more familiar and frequent users of GAI tools than instructors}

The majority of the students were either familiar with the capabilities of GAI tools or use them frequently. On the contrary, some instructors had tried GAI tools, but they did not use them very frequently.

\underline{\textit{Challenge \#1:}} Instructors need to become familiar with the functionality and limitations of GAI tools in order to adapt curricula and assessment methods adequately. Currently, instructors are at a disadvantage to students who are much more familiar with these models. 

\subsubsection{Students and instructors are not aligned on how to adapt assessment}

While both stakeholders expressed uncertainty about how to adapt assessment methods to account for GAI tools, the suggestions they did provide did not align. Students emphasized that the weight and frequency of hands-on, active learning assignments, such as programming labs, should not change as they are valuable to students’ comprehension of new material and learning goals. A few of the instructors proposed to adapt their assessment methods to combat academic dishonesty by reducing the weight of some of the lab assignments and increasing the weight or frequency of in-class, proctored assignments, such as quizzes or tests. This divergence between the values of students and instructors has the potential to create significant tension between students and instructors.

\underline{\textit{Challenge \#2:}} Instructors wanted to develop assessments that mitigate academic dishonesty. One-third of interviewed instructors suggested doing this by increasing the weight of proctored exams.
However, students expressed that they still want hands-on and active learning assignments (e.g., programming lab assignments) to represent a larger portion of their overall grade, because, as many students expressed, they learned better when they were engaged in the tasks they had to complete. 
Increasing the frequency or weight of in-class proctored exams moves away from evidence-based active learning~\cite{bonwell_active_1991}, additionally it may negatively affect students.

\subsubsection{Instructors were not aware of students’ motivations for using GAI tools}
Students shared a variety of reasons as to why they would choose to use GAI tools to assist with or complete their assignments, such as reducing busy work, getting another perspective other than their instructor on a concept, and getting help when their instructor is not available. When instructors were asked to consider why students use these models, the only common response was regarding students using GAI tools for busy work.

\underline{\textit{Challenge \#3:}} Instructors need to better understand students' motivations for using GAI tools. Without this understanding, instructors could potentially make changes to pedagogy and assessment that unintentionally harm students. To address this concern, researchers and practitioners could co-design solutions with students. Co-design is useful for curriculum design~\cite{gunckel_including_2005} and has already been used for integrating AI into classrooms~\cite{holstein_co-designing_2019}. These techniques may not only result in better, more equitable solutions but also have the potential to build relationships and leverage students' deeper familiarity with these GAI tools.

\subsection{Conclusion}
This paper presents the first systematic investigation of both students' and instructors' experiences with and preferences for using GAI tools in computing classrooms. While these results are preliminary, we believe that rapidly disseminating these critical views can guide researchers to capitalize on opportunities while mitigating potential challenges.  

\section*{Acknowledgment}

We would like to thank Dr. Brian McInnis for feedback on the abstract which helped refine the final framing of the paper. 

\bibliographystyle{IEEEtran}
\bibliography{references.bib}

\section*{Appendix}
\textbf{Semi-Structured Interview Script}\\
\textit{Personal Background Information:}
\begin{enumerate}
    \item (Students) What is your major? What year are you? What kinds of classes have you taken?
    \item (Instructors) What courses do you teach? How many years have you been teaching? What kinds of assignments do you to tend to have students complete?
\end{enumerate}
\textit{Awareness of GAI tools:}
\begin{enumerate}
    \item Are you familiar with GAI tools like GitHub CoPilot, Amazon CodWhisperer, ChatGPT, Grammarly, or Google Smart Compose?"
\end{enumerate}
\textit{Prior Experiences with GAI tools:}
\begin{enumerate}
    \item Have you ever used a GAI tool for an educational reason? If so, describe a time when you used a GAI tool to help with an assignment.
    \item What kind of assignment was it for? How was it helpful/unhelpful? Would you use it again? Why did you choose to use it?
\end{enumerate}
\textit{Values Held for GAI tools:}
\begin{enumerate}
    \item (Students) What types of coursework/assignments do you find yourself needing assistance on?
    \item (Students) When would you use a tool like these in addition to or instead of getting assistance from your instructor, teaching assistants, or peers?
    \item (Students) What types of assignments do you think are the most beneficial to your ability to comprehend new material?
    \item What situations do you think these tools could be beneficial for students? For instructors?
    \item What are your concerns for their usage by students? By instructors?
    \item How often do you believe students use tools like these to plagiarize work on assignments?
    \item What curriculum changes do you expect will be required as a result of these models?
\end{enumerate}

\end{document}